\documentclass[proof]{WileyASNA-v1}

\usepackage{xcolor}
\usepackage{soul}

\articletype{Original article}%
\received{28 Nov 2025}
\revised{20 Jan 2026}
\accepted{26 Jan 2026}

\begin{document}

\title{Searching for rotational X-ray modulation on TIC 277539431}

\author[1,2]{Desmond Dsouza*}

\author[1,2]{Katja Poppenh\"ager}

\author[3]{Ekaterina Ilin }

\authormark{Desmond Dsouza \textsc{et al}}

\address[1]{ \orgname{Leibniz Institute for Astrophysics Potsdam (AIP)}, \orgaddress{\state{Potsdam}, \country{Germany}}}

\address[2]{\orgdiv{Institute for Physics and Astronomy}, \orgname{Universit\"at Potsdam}, \orgaddress{\state{Potsdam-Golm}, \country{Germany}}}

\address[3]{\orgdiv{Netherlands Institute for Radio Astronomy}, \orgname{ASTRON}, \orgaddress{\state{Dwingeloo}, \country{Netherlands}}}

\corres{*Desmond Dsouza, Leibniz Institute for Astrophysics Potsdam (AIP), An der Sternwarte 16, 14482 Potsdam. \email{ddsouza@aip.de}}

\presentaddress{Leibniz Institute for Astrophysics Potsdam (AIP), An der Sternwarte 16, 14482 Potsdam.}

\abstract{TIC 277539431, a fast rotating M7 dwarf, was detected to host the highest latitude flare to date at $81^\circ$. Magnetic activity like stellar flares occurring at high latitude indicate occurrence of coronal loops at these latitudes on fully-convective M dwarfs. In contrast, sunspots usually occur below $30^\circ$. In our study we look for modulation on the X-ray signal occurring due to occultation of coronal loops by the star due to stellar rotation. We report an updated rotation period for this star as $P_{\text{rot}}=273.593$ min based on TESS sectors 12, 37, 39, 64 and 65. We conducted $\chi^2_{\textrm{red}}$ fits by varying the amplitude and the phase of a sinusoidally modulated signal derived from the new rotation period. We find no evidence of rotational modulation in the X-ray signal. This could be due to multiple scenarios, such as lack of a stable coronal loop during observation or the modulated signal being too weak, however given the dataset, individual scenarios cannot be distinguished.}

		
\jnlcitation{\cname{%
\author{D. Dsouza}, 
\author{K. Poppenh\"ager}, and \author{E. Ilin}}
(\cyear{2024}), 
\ctitle{Searching for rotational X-ray modulation on TIC 277539431}, \cvol{xxxx:xx-xx}.}


\maketitle

\section{Introduction}\label{sec1}
M dwarfs make up 70-75$\%$ of all stars in the solar neighbourhood \citep{2010AJ....139.2679B, 2019AJ....157..216W}. Due to their lower bolometric luminosity, their habitable zones lie closer to the host star. These conditions make them favourable targets to look for exoplanets and habitability studies \citep{2018A&A...609A.117T}. Contrary to the Sun, M-type stars, especially fully convective M dwarfs have much longer sustained saturated magnetic activity \citep{2022A&A...661A..29M} due to their long spindown timescales \citep{2022ApJ...936..109P}. As more exoplanets are discovered to be hosted by M dwarfs, studying the differences in magnetic activity becomes imperative	.	

Helioseismic studies have shown that the Sun is differentially rotating \citep{1984Natur.310...22D, 2000Sci...287.2456H}. This differential rotation leads to shearing regions called the tachocline that is believed to fuel the $\alpha \Omega$ dynamo which gives rise to magnetic fields on the Sun and sun-like partially convective stars \citep{1976A&A....47..243S, 2014ApJ...782...93H, 2014ARA&A..52..251C}. Although labelled as cool stars, fully convective M-type stars are believed to host a different dynamo mechanism than the $\alpha \Omega$ dynamo due to the missing tachocline region on such stars \citep{2006A&A...446.1027C, 2015ApJ...813L..31Y, 2016ApJ...833L..28Y, 2018ApJ...859...18P}.

Observations of fully convective M dwarfs show that they display magnetic activity like that of other cool stars. Coronal X-ray emission \citep{2018MNRAS.479.2351W, 2022A&A...661A..29M}, chromospheric activity indicators like Ca II H$\&$K, H$\alpha$ \citep{2008ApJ...684.1390R, 2017ApJ...834...85N}, flares \citep{2021MNRAS.507.1723I} and other magnetic activity are frequently observed on these stars \citep{2021A&ARv..29....1K}. 

\cite{2021MNRAS.507.1723I} observed that white-light flares on fast-rotating fully convective M dwarfs predominantly occur at high latitudes. TIC 277539431 was observed to have a high latitude flare at $81^\circ$ in their study. In contrast, the Sun hosts most of its activity such as sunspots at or lower than $\approx 30^\circ$. Since flares occur due to re-connection of magnetic flux ropes \citep{2010ARA&A..48..241B}, the high latitude flare on TIC 277539431 indicates the existence of flux loops at such latitudes. 

Our study looks at TIC 277539431 using the X-ray telescope \textit{XMM-Newton} to look for modulation in the coronal emission. If coronal loops occur at such high latitudes, we expect them to be partially occulted during stellar rotation, causing a modulated X-ray signal. Rotational X-ray modulation is observed on the Sun \citep{2004A&ARv..12...71G}, however, it has been rarely observed on other cool stars. The exceptions being EK Dra \citep{1995A&A...301..201G} and IC 2391 \citep{2003A&A...407L..63M}. Detections have been challenging due to the nature of X-ray observations, which typically involve low photon count rates and therefore large errorbars compared to the expected amplitude of such a modulation, making detections of X-ray modulation with short baselines difficult.

A measurable modulation in the X-ray signal could, however, allow us to directly measure the coronal loop heights on this star. While scaling relations from X-ray spectra allows for estimation of loop size and Magnetic field on TIC 277539431 \citep{2024A&A...687A.138I}, a measurement of loop height through modulation of X-ray signal could offer an empherical method to study the corona of fully-convective stars and their surroundings for habitability studies. The feasibility of detecting an X-ray modulation, however, depends on the occulted volume of the loop; i.e., a large coronal loop on the pole that is only barely occulted could cause a very small modulation of the X-ray lightcurve.

In the next section we describe the observation and data analysis conducted in our study. In section 3, we discuss our results on how photometric modulation observed in TESS was used to look for the X-ray modulation, then in section 4 and 5 we discuss and conclude our findings.

\begin{center}
\begin{table}[t]%
\centering
\caption{Stellar parameters of TIC 277539431.\label{tab4}}%
\begin{tabular*}{20pc}{@{\extracolsep\fill}lcc@{\extracolsep\fill}}%
\toprule
\textbf{Parameter} & \textbf{Value}  \\
\midrule
Spectral type & M7 \tnote{$\dagger$}  \\
Distance d & 13.70$\pm 0.11$ pc \tnote{$\ddagger$}\\
Effective temperature T$_{\text{eff}}$ & $2680_{-50}^{60}$K \tnote{$\S$}\\
Rotation period P$_{\text{rot}}$ & 273.593 min \tnote{\P}\\
Projected rot. vel $v\sin i$ & 38.6 $\pm 1.0\text{kms}^{-1}$ \tnote{$\dagger$}\\
Inclination i & $(87.0 ^{+2.0}_{-2.4})^\circ$ \tnote{$\dagger$}\\
Radius R & $0.145\pm 0.004 \text{R}_\odot$ \tnote{$\dagger$} \\

\bottomrule
\end{tabular*}
\begin{tablenotes}

Reference: 
\item[$\dagger$] \cite{2021A&A...645A..42I}
\item[$\ddagger$] \cite{2018AJ....156...58B}
\item[$\S$ ] \cite{2013ApJS..208....9P}
\item[$\P$ ] This work
\end{tablenotes}
\end{table}
\end{center}

\section{Observations and data analysis}\label{sec2}
\begin{figure}[t]
	\centerline{\includegraphics[width=250pt]{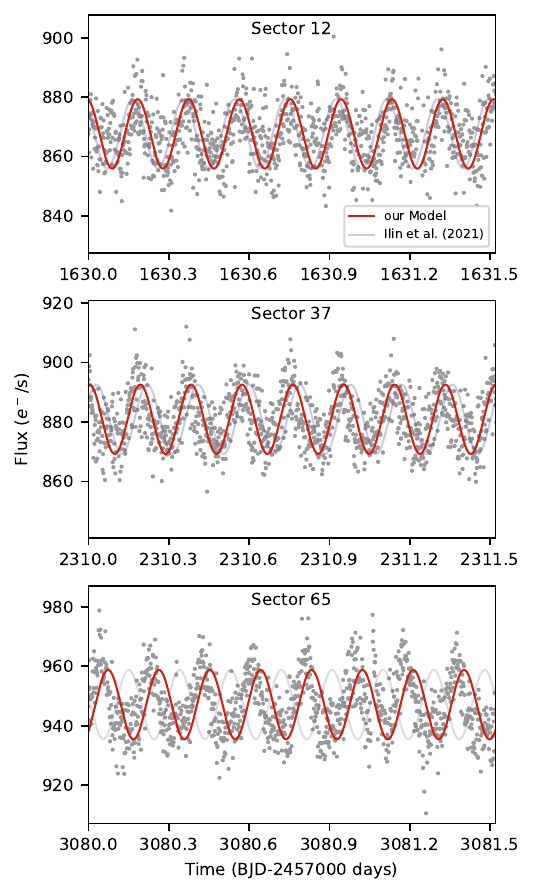}}
	\caption{Selected excerpts of TESS (\texttt{PDCSAP\_FLUX}) lightcurves  from sectors 12, 37 and 65. Our model using the $P_{\text{rot}}=273.593$ min matches the modulation in all TESS sectors in contrast to the previously reported $P_{\text{rot}} = 273.618\pm 0.007$ min. \label{fig: TESS LC}}
\end{figure}

Proprietary observing time was granted to PI: Ekaterina Ilin (author 3) to observe TIC 277539431 on the XMM-Newton, X-ray space observatory for 36ks ($\approx 2.2P_{\text{rot}}$). The observation took place on August 5, 2022. The XMM-Newton telescope \citep{2001A&A...365L...1J} was launched by the European Space Agency on December 1999. It can observe a patch of the sky simultaneously with 3 X-ray detectors: MOS1, MOS2 and PN (\cite{2001A&A...365L..27T} and \cite{2001A&A...365L..18S}), Additionally the space observatory also hosts an Optical Monitor (OM).

We used the XMM-Newton Science Analysis System (SAS) version 21.0.0 to process the X-ray data. The X-ray photons from the source and the background regions were extracted using the SAS command \texttt{evselect} with a time bin of 407s (90 bins). The light-curve was then generated using \texttt{epiclccorr} with absolute correction and background correction modes turned on. We additionally identified the flare as done by \cite{2024A&A...687A.138I}. A flare occurs when magnetic flux ropes combine or are disrupted \citep{2010ARA&A..48..241B}. Since our study searches for X-ray modulation occuring due to stable flux loops near the pole of the star, we removed the flare from the analysis. We expect stable polar flux ropes to come in and out of view due to stellar rotation, however the stability of the flux ropes could have been lost during the occurance of a flare (if the flare occured near the pole). On the other hand, if the flare was not of polar origin then it would add uncertainity in our analysis. This is why we remove the flaring part of the lightcurve for our study. As noted in \cite{2024A&A...687A.138I}, TIC 277539431 was observed with the OM, however the low signal to noise ratio shows no optical modulation, this is why we resort to the data from the Transiting Exoplanet Survey Satellite (TESS) in our study.

TESS observed TIC 277539431 during observing sectors 12, 37, 39, 64 and 65. These sectors span 4 years from mid 2019 to mid 2023. Although TESS was launched by National Aeronautics and Space Administration (NASA) to primarily detect exoplanets around bright stars \citep{2015JATIS...1a4003R}, its photometric cadence of 2 minutes is useful for detecting rotation periods ($\text{P}_{\text{rot}}$) of fast rotating stars. The online tool by the Mikulski archive for space telescopes was used to retrieve the TESS lightcurves (LC) for our study. TIC 277539431 was previously reported with a $\text{P}_{\text{rot}}=273.618\pm 0.007$ min \citep{2021MNRAS.507.1723I}. 

\section{Results}\label{Results} 

\subsection{Photometric modulation}\label{photometric modulation}
We observe that the modulated signal in the TESS lightcurves previously reported by \cite{2021MNRAS.507.1723I} can also be seen in sectors observed later, i.e. 37, 39, 64 and 65. However, the estimated $P_{\text{rot}} = 273.618\pm 0.007$ min is a good fit for only sector 12 and 37 (Fig.\ref{fig: TESS LC}). A Lomb-Scargle periodogram \citep{1976Ap&SS..39..447L, 1982ApJ...263..835S} taking into account all sectors give us a peak at $P_{\text{rot}}=273.593$ min.  We use the False Alarm Probability (FAP) to check reliability of this peak. The inbuilt astropy functions \citep{2022ApJ...935..167A} for the Baluev method \citep{2008MNRAS.385.1279B} and the Bootstrap method both give us a FAP of $\le 10^{-10}$. We found that this slightly different period is a better match for all the observed TESS sectors up to and including sector 65 (As seen in Fig.\ref{fig: TESS LC}). 

The amplitude of the modelled LC was calculated using the averaged standard deviations $\sigma$ of individual sectors. Note that the amplitude of the TESS signal is not relevant for our study. It has been calculated only to visually compare the $P_{\text{rot}}$ estimated in this study with respect to that reported previously. 

\subsection{Searching for rotational X-ray modulation}\label{sec3.2}

\begin{figure}[t]
	\centerline{\includegraphics[width=250pt]{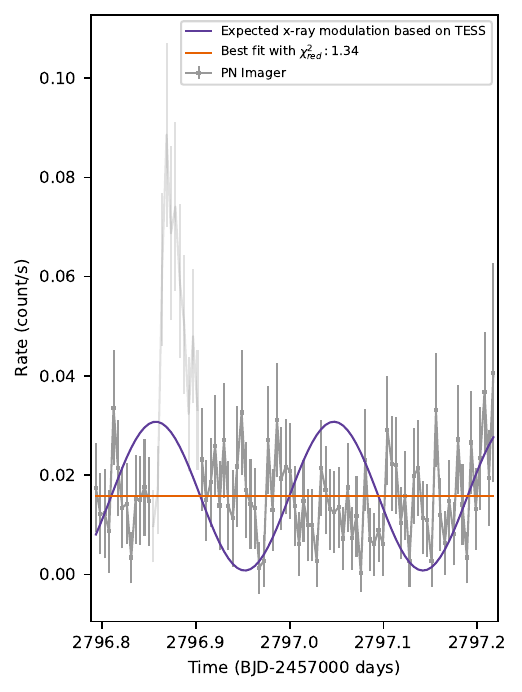}}
	\caption{X-ray lightcurve data from the PN detector. Light gray area represents the removed flare. Purple curve is representative of the modulating signal we expect in X-ray based on the period estimated from TESS ($180^\circ$ phase shift). Orange curve represents the best $\chi^2_{\text{red}}$ fit, i.e a straight line. \label{XMM LC with fits plot}}
\end{figure}

We modelled the X-ray lightcurve with a grid of sinusoidally modulated emission using different amplitudes and phase shifts with respect to the observed photospheric modulation seen in TESS. Our model is based on the standard wave equation (Eq. \ref{flux model equation}).{ }Here, $f_\text{model}(t)$ is the modelled flux dependent on time t, $P_{\text{rot}}$ is the rotation period obtained in Sec. \ref{photometric modulation} and $\mu$ is the average observed flux obtained from the X-ray lightcurve excluding the flare. The two free parameters are amplitude $A$ and phase $\phi$. 

\begin{equation}
	f_\text{model}(t)=A \sin{\Big( \frac{2\pi t}{P_{\text{rot}}} + \phi\Big)} + \mu_{}
 \label{flux model equation}
\end{equation}

The wave models are generated by varying the free parameters. A range of amplitudes from 0 count/s (flat line) to 0.0175 count/s with a step size of 0.0025 count/s and phases from $0^\circ$ to $330^\circ$ with step size of $30^\circ$ are probed.

To find the best fit, we employ the reduced $\chi^2_{\text{red}}$ test. The degree of freedom is the number of observations (90 bins) subtracted by the number of fitting parameters. In our case, we have 2 fitting parameters, that is A and $\phi$. However, for the case of $A= 0.0$ count/s (straight line), we have zero fitting parameters.

\begin{figure}[t]
	\centerline{\includegraphics[width=250pt]{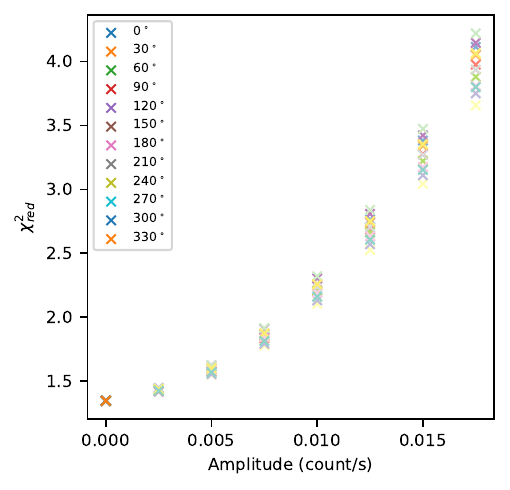}}
	\caption{Reduced chi-square fit ($ \chi^2_{\text{red}}$) of the X-ray observation to the modelled grid of varying fluxes. The lower the $ \chi^2_{\text{red}}$ value, the better the fit. The best fit ($ \chi^2_{\text{red}} = 1.34$) corresponds to a wave with zero amplitude, i.e a straight line. \label{Chi2 fit plot}}
\end{figure}

The best fit is when $\chi^2_{\text{red}}$ is minimised to $\approx1$. Figure \ref{Chi2 fit plot} shows the best fit to our data is $\chi^2_{\text{red}}$ = 1.34. This corresponds to the model of a straight line. Normally $\chi^2_{\text{red}}$ $\ge1$ indicates an under-fit, however, from the grid of simulated wave models, this was found to be the best fit to our dataset. In Figure \ref{XMM LC with fits plot}we plot the PN lightcurve along with the expected X-ray modulation we would get if the TESS modelled lightcurve was phase shifted by $180^\circ$. We also plot the best fit to our data.

\section{Discussion}
Observations of the Sun show that sun spots appear darker on the photosphere. Sun spots are sources of magnetic flux ropes which appear bright in X-ray and UV due to the energy it dissipates into the chromosphere and the corona. Therefore, a detectable X-ray modulation would be at anti-correlation to the photometric modulation. This is expected to be a phase ($\phi$) shift of $180^\circ$. However, the best $\chi^2_{\text{red}}$ fit was found to be a straight line. In this section we look at the possible scenarios for this result.

\textit{Faint source}: M7 dwarfs, although persistently magnetically active for Gigayears, tend to be relatively dim in X-rays due to their small surface area. Therefore, our coverage of only $P_{\textrm{rot}}$ may have been too short to detect a recurring modulation at the present noise level in the data.

\textit{No stable coronal loop during observation}: Coronal loops are a transient phenomenon, their appearance and disappearance for stars cannot be predicted. During our observation, there may have been no high latitude Coronal loops. There is also the possibility that any existing stable loop was disrupted during the flaring event at the time of observation.

\textit{Modulated signal is weak.}: M dwarfs are known to host massive coronal loops, sometimes much bigger than their stellar radius \citep{2005A&A...436.1041M}. Assuming there was a massive coronal loop at a high latitude, a partial occultation of this coronal flux rope due to stellar rotation could reduce the X-ray luminosity minimally. This small change in signal could be washed out by the intrinsic variability of the quiescent Corona. A weak modulated signal with an amplitude of 0.018 ct/s or less could be masked in the signal due to the RMS fluctuation of the data. It should be noted that the flare (Greyed out in Fig \ref{XMM LC with fits plot}) was previously reported to have a loop length between 0.02 - 0.13 $R_*$ \citep{2024A&A...687A.138I}. However, it is unclear at which latitude this particular flare originates from. 

\textit{Other activity sources during the same time}: TIC 277539431, due to its fast rotation, lies deep in the coronal saturated regime \citep{2018MNRAS.479.2351W}. This indicates that the corona is saturated with coronal activity, therefore the variability of the corona in the other parts of the star could be stronger than the modulation we expect to see for an occulted coronal loop.

The scenarios mentioned above could be the reason for a non-detection. Given the dataset however, the effect of the individual scenarios cannot be distinguished.

\section{Conclusion}\label{sec4}
TIC 277539431 was observed in optical (TESS) and X-ray (XMM-Newton) wavelengths. Our goal was to test observational feasibility for modulation on the X-ray signal caused due to partial occultation of high latitude coronal loops. Due to the newly available TESS data, we could constrain the rotation period of the star to be $P_{\text{rot}}=273.593$ min for TESS sectors 12, 37, 39, 64 and 65. However, no evidence of rotational modulation of the stellar X-ray emission could be found, indicating a similar X-ray brightness across stellar longitudes at the time of observation. 

Future space missions such as the European Space Agency’s  \textit{NewAthena} promises to increase the collecting area by an order of magnitude when compared to the XMM-Newton \citep{2025NatAs...9...36C}. After its launch, \textit{NewAthena} could enable studies of X-ray modulation for later type stars.

\section*{Acknowledgments}
The authors wish to thank the anonymous referee for their useful suggestions and comments. The authors thank Dr. Nikoleta Ilić and Dr. Julián Alvarado-Gómez for their discussion that helped enriched this paper. This work was funded by \fundingAgency{DLR} under funding number \fundingNumber{50 OR 2209}. This work made use of Astropy:\footnote{http://www.astropy.org} a community-developed core Python package and an ecosystem of tools and resources for astronomy \citep{2013A&A...558A..33A, 2018AJ....156..123A, 2022ApJ...935..167A}.

\appendix

\bibliography{Wiley-ASNA}%

@ARTICLE{1976A&A....47..243S,
       author = {{Stix}, M.},
        title = "{Differential rotation and the solar dynamo.}",
      journal = {\aap},
         year = 1976,
        month = mar,
       volume = {47},
       number = {2},
        pages = {243-254},
       adsurl = {https://ui.adsabs.harvard.edu/abs/1976A&A....47..243S}
}

@ARTICLE{1976Ap&SS..39..447L,
       author = {{Lomb}, N.~R.},
        title = "{Least-Squares Frequency Analysis of Unequally Spaced Data}",
      journal = {\apss},
         year = 1976,
        month = feb,
       volume = {39},
       number = {2},
        pages = {447-462},
          doi = {10.1007/BF00648343},
       adsurl = {https://ui.adsabs.harvard.edu/abs/1976Ap&SS..39..447L}
}

@ARTICLE{1982ApJ...263..835S,
       author = {{Scargle}, J.~D.},
        title = "{Studies in astronomical time series analysis. II. Statistical aspects of spectral analysis of unevenly spaced data.}",
      journal = {\apj},
         year = 1982,
        month = dec,
       volume = {263},
        pages = {835-853},
          doi = {10.1086/160554},
       adsurl = {https://ui.adsabs.harvard.edu/abs/1982ApJ...263..835S}
}

@ARTICLE{1984Natur.310...22D,
       author = {{Duvall}, T.~L., Jr. and {Dziembowski}, W.~A. and {Goode}, P.~R. and {Gough}, D.~O. and {Harvey}, J.~W. and {Leibacher}, J.~W.},
        title = "{Internal rotation of the Sun}",
      journal = {\nat},
         year = 1984,
        month = jul,
       volume = {310},
       number = {5972},
        pages = {22-25},
          doi = {10.1038/310022a0},
       adsurl = {https://ui.adsabs.harvard.edu/abs/1984Natur.310...22D}
}

@ARTICLE{1995A&A...301..201G,
       author = {{Guedel}, M. and {Schmitt}, J.~H.~M.~M. and {Benz}, A.~O. and {Elias}, II, N.~M.},
        title = "{The corona of the young solar analog EK Draconis.}",
      journal = {\aap},
         year = 1995,
        month = sep,
       volume = {301},
        pages = {201},
       adsurl = {https://ui.adsabs.harvard.edu/abs/1995A&A...301..201G}
}

@ARTICLE{2000Sci...287.2456H,
       author = {{Howe}, R. and {Christensen-Dalsgaard}, J. and {Hill}, F. and {Komm}, R.~W. and {Larsen}, R.~M. and {Schou}, J. and {Thompson}, M.~J. and {Toomre}, J.},
        title = "{Dynamic Variations at the Base of the Solar Convection Zone}",
      journal = {Science},
         year = 2000,
        month = mar,
       volume = {287},
       number = {5462},
        pages = {2456-2460},
          doi = {10.1126/science.287.5462.2456},
       adsurl = {https://ui.adsabs.harvard.edu/abs/2000Sci...287.2456H}
}

@ARTICLE{2001A&A...365L...1J,
       author = {{Jansen}, F. and {Lumb}, D. and {Altieri}, B. and {Clavel}, J. and {Ehle}, M. and {Erd}, C. and {Gabriel}, C. and {Guainazzi}, M. and {Gondoin}, P. and {Much}, R. and {Munoz}, R. and {Santos}, M. and {Schartel}, N. and {Texier}, D. and {Vacanti}, G.},
        title = "{XMM-Newton observatory. I. The spacecraft and operations}",
      journal = {\aap},
         year = 2001,
        month = jan,
       volume = {365},
        pages = {L1-L6},
          doi = {10.1051/0004-6361:20000036},
       adsurl = {https://ui.adsabs.harvard.edu/abs/2001A&A...365L...1J}
}

@ARTICLE{2001A&A...365L..18S,
       author = {{Str{\"u}der}, L. and {Briel}, U. and {Dennerl}, K. and {Hartmann}, R. and {Kendziorra}, E. and {Meidinger}, N. and {Pfeffermann}, E. and {Reppin}, C. and {Aschenbach}, B. and {Bornemann}, W. and {Br{\"a}uninger}, H. and {Burkert}, W. and {Elender}, M. and {Freyberg}, M. and {Haberl}, F. and {Hartner}, G. and {Heuschmann}, F. and {Hippmann}, H. and {Kastelic}, E. and {Kemmer}, S. and {Kettenring}, G. and {Kink}, W. and {Krause}, N. and {M{\"u}ller}, S. and {Oppitz}, A. and {Pietsch}, W. and {Popp}, M. and {Predehl}, P. and {Read}, A. and {Stephan}, K.~H. and {St{\"o}tter}, D. and {Tr{\"u}mper}, J. and {Holl}, P. and {Kemmer}, J. and {Soltau}, H. and {St{\"o}tter}, R. and {Weber}, U. and {Weichert}, U. and {von Zanthier}, C. and {Carathanassis}, D. and {Lutz}, G. and {Richter}, R.~H. and {Solc}, P. and {B{\"o}ttcher}, H. and {Kuster}, M. and {Staubert}, R. and {Abbey}, A. and {Holland}, A. and {Turner}, M. and {Balasini}, M. and {Bignami}, G.~F. and {La Palombara}, N. and {Villa}, G. and {Buttler}, W. and {Gianini}, F. and {Lain{\'e}}, R. and {Lumb}, D. and {Dhez}, P.},
        title = "{The European Photon Imaging Camera on XMM-Newton: The pn-CCD camera}",
      journal = {\aap},
         year = 2001,
        month = jan,
       volume = {365},
        pages = {L18-L26},
          doi = {10.1051/0004-6361:20000066},
       adsurl = {https://ui.adsabs.harvard.edu/abs/2001A&A...365L..18S}
}

@ARTICLE{2001A&A...365L..27T,
       author = {{Turner}, M.~J.~L. and {Abbey}, A. and {Arnaud}, M. and {Balasini}, M. and {Barbera}, M. and {Belsole}, E. and {Bennie}, P.~J. and {Bernard}, J.~P. and {Bignami}, G.~F. and {Boer}, M. and {Briel}, U. and {Butler}, I. and {Cara}, C. and {Chabaud}, C. and {Cole}, R. and {Collura}, A. and {Conte}, M. and {Cros}, A. and {Denby}, M. and {Dhez}, P. and {Di Coco}, G. and {Dowson}, J. and {Ferrando}, P. and {Ghizzardi}, S. and {Gianotti}, F. and {Goodall}, C.~V. and {Gretton}, L. and {Griffiths}, R.~G. and {Hainaut}, O. and {Hochedez}, J.~F. and {Holland}, A.~D. and {Jourdain}, E. and {Kendziorra}, E. and {Lagostina}, A. and {Laine}, R. and {La Palombara}, N. and {Lortholary}, M. and {Lumb}, D. and {Marty}, P. and {Molendi}, S. and {Pigot}, C. and {Poindron}, E. and {Pounds}, K.~A. and {Reeves}, J.~N. and {Reppin}, C. and {Rothenflug}, R. and {Salvetat}, P. and {Sauvageot}, J.~L. and {Schmitt}, D. and {Sembay}, S. and {Short}, A.~D.~T. and {Spragg}, J. and {Stephen}, J. and {Str{\"u}der}, L. and {Tiengo}, A. and {Trifoglio}, M. and {Tr{\"u}mper}, J. and {Vercellone}, S. and {Vigroux}, L. and {Villa}, G. and {Ward}, M.~J. and {Whitehead}, S. and {Zonca}, E.},
        title = "{The European Photon Imaging Camera on XMM-Newton: The MOS cameras}",
      journal = {\aap},
         year = 2001,
        month = jan,
       volume = {365},
        pages = {L27-L35},
          doi = {10.1051/0004-6361:20000087},
archivePrefix = {arXiv},
       eprint = {astro-ph/0011498},
 primaryClass = {astro-ph},
       adsurl = {https://ui.adsabs.harvard.edu/abs/2001A&A...365L..27T}
}

@ARTICLE{2003A&A...407L..63M,
       author = {{Marino}, A. and {Micela}, G. and {Peres}, G. and {Sciortino}, S.},
        title = "{X-ray rotational modulation of a supersaturated star in IC 2391}",
      journal = {\aap},
         year = 2003,
        month = aug,
       volume = {407},
        pages = {L63-L66},
          doi = {10.1051/0004-6361:20031053},
archivePrefix = {arXiv},
       eprint = {astro-ph/0307170},
 primaryClass = {astro-ph},
       adsurl = {https://ui.adsabs.harvard.edu/abs/2003A&A...407L..63M}
}

@ARTICLE{2004A&ARv..12...71G,
       author = {{G{\"u}del}, Manuel},
        title = "{X-ray astronomy of stellar coronae}",
      journal = {\aapr},
         year = 2004,
        month = sep,
       volume = {12},
       number = {2-3},
        pages = {71-237},
          doi = {10.1007/s00159-004-0023-2},
archivePrefix = {arXiv},
       eprint = {astro-ph/0406661},
 primaryClass = {astro-ph},
       adsurl = {https://ui.adsabs.harvard.edu/abs/2004A&ARv..12...71G}
}

@ARTICLE{2005A&A...436.1041M,
       author = {{Mitra-Kraev}, U. and {Harra}, L.~K. and {Williams}, D.~R. and {Kraev}, E.},
        title = "{The first observed stellar X-ray flare oscillation: Constraints on the flare loop length and the magnetic field}",
      journal = {\aap},
         year = 2005,
        month = jun,
       volume = {436},
       number = {3},
        pages = {1041-1047},
          doi = {10.1051/0004-6361:20052834},
archivePrefix = {arXiv},
       eprint = {astro-ph/0503384},
 primaryClass = {astro-ph},
       adsurl = {https://ui.adsabs.harvard.edu/abs/2005A&A...436.1041M}
}

@ARTICLE{2006A&A...446.1027C,
       author = {{Chabrier}, G. and {K{\"u}ker}, M.},
        title = "{Large-scale {\ensuremath{\alpha}}\^2-dynamo in low-mass stars and brown dwarfs}",
      journal = {\aap},
         year = 2006,
        month = feb,
       volume = {446},
       number = {3},
        pages = {1027-1037},
          doi = {10.1051/0004-6361:20042475},
archivePrefix = {arXiv},
       eprint = {astro-ph/0510075},
 primaryClass = {astro-ph},
       adsurl = {https://ui.adsabs.harvard.edu/abs/2006A&A...446.1027C}
}

@ARTICLE{2008ApJ...684.1390R,
       author = {{Reiners}, A. and {Basri}, G.},
        title = "{Chromospheric Activity, Rotation, and Rotational Braking in M and L Dwarfs}",
      journal = {\apj},
         year = 2008,
        month = sep,
       volume = {684},
       number = {2},
        pages = {1390-1403},
          doi = {10.1086/590073},
archivePrefix = {arXiv},
       eprint = {0805.1059},
 primaryClass = {astro-ph},
       adsurl = {https://ui.adsabs.harvard.edu/abs/2008ApJ...684.1390R}
}

@ARTICLE{2008MNRAS.385.1279B,
       author = {{Baluev}, R.~V.},
        title = "{Assessing the statistical significance of periodogram peaks}",
      journal = {\mnras},
         year = 2008,
        month = apr,
       volume = {385},
       number = {3},
        pages = {1279-1285},
          doi = {10.1111/j.1365-2966.2008.12689.x},
archivePrefix = {arXiv},
       eprint = {0711.0330},
 primaryClass = {astro-ph},
       adsurl = {https://ui.adsabs.harvard.edu/abs/2008MNRAS.385.1279B}
}

@ARTICLE{2010AJ....139.2679B,
       author = {{Bochanski}, John J. and {Hawley}, Suzanne L. and {Covey}, Kevin R. and {West}, Andrew A. and {Reid}, I. Neill and {Golimowski}, David A. and {Ivezi{\'c}}, {\v{Z}}eljko},
        title = "{The Luminosity and Mass Functions of Low-mass Stars in the Galactic Disk. II. The Field}",
      journal = {\aj},
         year = 2010,
        month = jun,
       volume = {139},
       number = {6},
        pages = {2679-2699},
          doi = {10.1088/0004-6256/139/6/2679},
archivePrefix = {arXiv},
       eprint = {1004.4002},
 primaryClass = {astro-ph.SR},
       adsurl = {https://ui.adsabs.harvard.edu/abs/2010AJ....139.2679B}
}

@ARTICLE{2010ARA&A..48..241B,
       author = {{Benz}, Arnold O. and {G{\"u}del}, Manuel},
        title = "{Physical Processes in Magnetically Driven Flares on the Sun, Stars, and Young Stellar Objects}",
      journal = {\araa},
         year = 2010,
        month = sep,
       volume = {48},
        pages = {241-287},
          doi = {10.1146/annurev-astro-082708-101757},
       adsurl = {https://ui.adsabs.harvard.edu/abs/2010ARA&A..48..241B}
}

@ARTICLE{2013ApJS..208....9P,
       author = {{Pecaut}, Mark J. and {Mamajek}, Eric E.},
        title = "{Intrinsic Colors, Temperatures, and Bolometric Corrections of Pre-main-sequence Stars}",
      journal = {\apjs},
         year = 2013,
        month = sep,
       volume = {208},
       number = {1},
          eid = {9},
        pages = {9},
          doi = {10.1088/0067-0049/208/1/9},
archivePrefix = {arXiv},
       eprint = {1307.2657},
 primaryClass = {astro-ph.SR},
       adsurl = {https://ui.adsabs.harvard.edu/abs/2013ApJS..208....9P}
}

@ARTICLE{2013A&A...558A..33A,
       author = {{Astropy Collaboration} and {Robitaille}, Thomas P. and {Tollerud}, Erik J. and {Greenfield}, Perry and {Droettboom}, Michael and {Bray}, Erik and {Aldcroft}, Tom and {Davis}, Matt and {Ginsburg}, Adam and {Price-Whelan}, Adrian M. and {Kerzendorf}, Wolfgang E. and {Conley}, Alexander and {Crighton}, Neil and {Barbary}, Kyle and {Muna}, Demitri and {Ferguson}, Henry and {Grollier}, Fr{\'e}d{\'e}ric and {Parikh}, Madhura M. and {Nair}, Prasanth H. and {Unther}, Hans M. and {Deil}, Christoph and {Woillez}, Julien and {Conseil}, Simon and {Kramer}, Roban and {Turner}, James E.~H. and {Singer}, Leo and {Fox}, Ryan and {Weaver}, Benjamin A. and {Zabalza}, Victor and {Edwards}, Zachary I. and {Azalee Bostroem}, K. and {Burke}, D.~J. and {Casey}, Andrew R. and {Crawford}, Steven M. and {Dencheva}, Nadia and {Ely}, Justin and {Jenness}, Tim and {Labrie}, Kathleen and {Lim}, Pey Lian and {Pierfederici}, Francesco and {Pontzen}, Andrew and {Ptak}, Andy and {Refsdal}, Brian and {Servillat}, Mathieu and {Streicher}, Ole},
        title = "{Astropy: A community Python package for astronomy}",
      journal = {\aap},
         year = 2013,
        month = oct,
       volume = {558},
          eid = {A33},
        pages = {A33},
          doi = {10.1051/0004-6361/201322068},
archivePrefix = {arXiv},
       eprint = {1307.6212},
 primaryClass = {astro-ph.IM},
       adsurl = {https://ui.adsabs.harvard.edu/abs/2013A&A...558A..33A}
}

@ARTICLE{2014ApJ...782...93H,
       author = {{Hazra}, Gopal and {Karak}, Bidya Binay and {Choudhuri}, Arnab Rai},
        title = "{Is a Deep One-cell Meridional Circulation Essential for the Flux Transport Solar Dynamo?}",
      journal = {\apj},
         year = 2014,
        month = feb,
       volume = {782},
       number = {2},
          eid = {93},
        pages = {93},
          doi = {10.1088/0004-637X/782/2/93},
archivePrefix = {arXiv},
       eprint = {1309.2838},
 primaryClass = {astro-ph.SR},
       adsurl = {https://ui.adsabs.harvard.edu/abs/2014ApJ...782...93H}
}

@ARTICLE{2014ARA&A..52..251C,
       author = {{Charbonneau}, Paul},
        title = "{Solar Dynamo Theory}",
      journal = {\araa},
         year = 2014,
        month = aug,
       volume = {52},
        pages = {251-290},
          doi = {10.1146/annurev-astro-081913-040012},
       adsurl = {https://ui.adsabs.harvard.edu/abs/2014ARA&A..52..251C}
}

@ARTICLE{2015JATIS...1a4003R,
       author = {{Ricker}, George R. and {Winn}, Joshua N. and {Vanderspek}, Roland and {Latham}, David W. and {Bakos}, G{\'a}sp{\'a}r {\'A}. and {Bean}, Jacob L. and {Berta-Thompson}, Zachory K. and {Brown}, Timothy M. and {Buchhave}, Lars and {Butler}, Nathaniel R. and {Butler}, R. Paul and {Chaplin}, William J. and {Charbonneau}, David and {Christensen-Dalsgaard}, J{\o}rgen and {Clampin}, Mark and {Deming}, Drake and {Doty}, John and {De Lee}, Nathan and {Dressing}, Courtney and {Dunham}, Edward W. and {Endl}, Michael and {Fressin}, Francois and {Ge}, Jian and {Henning}, Thomas and {Holman}, Matthew J. and {Howard}, Andrew W. and {Ida}, Shigeru and {Jenkins}, Jon M. and {Jernigan}, Garrett and {Johnson}, John Asher and {Kaltenegger}, Lisa and {Kawai}, Nobuyuki and {Kjeldsen}, Hans and {Laughlin}, Gregory and {Levine}, Alan M. and {Lin}, Douglas and {Lissauer}, Jack J. and {MacQueen}, Phillip and {Marcy}, Geoffrey and {McCullough}, Peter R. and {Morton}, Timothy D. and {Narita}, Norio and {Paegert}, Martin and {Palle}, Enric and {Pepe}, Francesco and {Pepper}, Joshua and {Quirrenbach}, Andreas and {Rinehart}, Stephen A. and {Sasselov}, Dimitar and {Sato}, Bun'ei and {Seager}, Sara and {Sozzetti}, Alessandro and {Stassun}, Keivan G. and {Sullivan}, Peter and {Szentgyorgyi}, Andrew and {Torres}, Guillermo and {Udry}, Stephane and {Villasenor}, Joel},
        title = "{Transiting Exoplanet Survey Satellite (TESS)}",
      journal = {Journal of Astronomical Telescopes, Instruments, and Systems},
         year = 2015,
        month = jan,
       volume = {1},
          eid = {014003},
        pages = {014003},
          doi = {10.1117/1.JATIS.1.1.014003},
       adsurl = {https://ui.adsabs.harvard.edu/abs/2015JATIS...1a4003R}
}

@ARTICLE{2015ApJ...813L..31Y,
       author = {{Yadav}, Rakesh K. and {Christensen}, Ulrich R. and {Morin}, Julien and {Gastine}, Thomas and {Reiners}, Ansgar and {Poppenhaeger}, Katja and {Wolk}, Scott J.},
        title = "{Explaining the Coexistence of Large-scale and Small-scale Magnetic Fields in Fully Convective Stars}",
      journal = {\apjl},
         year = 2015,
        month = nov,
       volume = {813},
       number = {2},
          eid = {L31},
        pages = {L31},
          doi = {10.1088/2041-8205/813/2/L31},
archivePrefix = {arXiv},
       eprint = {1510.05541},
 primaryClass = {astro-ph.SR},
       adsurl = {https://ui.adsabs.harvard.edu/abs/2015ApJ...813L..31Y}
}

@ARTICLE{2016ApJ...833L..28Y,
       author = {{Yadav}, Rakesh K. and {Christensen}, Ulrich R. and {Wolk}, Scott J. and {Poppenhaeger}, Katja},
        title = "{Magnetic Cycles in a Dynamo Simulation of Fully Convective M-star Proxima Centauri}",
      journal = {\apjl},
         year = 2016,
        month = dec,
       volume = {833},
       number = {2},
          eid = {L28},
        pages = {L28},
          doi = {10.3847/2041-8213/833/2/L28},
archivePrefix = {arXiv},
       eprint = {1610.02721},
 primaryClass = {astro-ph.SR},
       adsurl = {https://ui.adsabs.harvard.edu/abs/2016ApJ...833L..28Y}
}

@ARTICLE{2017ApJ...834...85N,
       author = {{Newton}, Elisabeth R. and {Irwin}, Jonathan and {Charbonneau}, David and {Berlind}, Perry and {Calkins}, Michael L. and {Mink}, Jessica},
        title = "{The H{\ensuremath{\alpha}} Emission of Nearby M Dwarfs and its Relation to Stellar Rotation}",
      journal = {\apj},
         year = 2017,
        month = jan,
       volume = {834},
       number = {1},
          eid = {85},
        pages = {85},
          doi = {10.3847/1538-4357/834/1/85},
archivePrefix = {arXiv},
       eprint = {1611.03509},
 primaryClass = {astro-ph.SR},
       adsurl = {https://ui.adsabs.harvard.edu/abs/2017ApJ...834...85N}
}

@ARTICLE{2018ApJ...859...18P,
       author = {{Pipin}, V.~V. and {Yokoi}, N.},
        title = "{Generation of a Large-scale Magnetic Field in a Convective Full-sphere Cross-helicity Dynamo}",
      journal = {\apj},
         year = 2018,
        month = may,
       volume = {859},
       number = {1},
          eid = {18},
        pages = {18},
          doi = {10.3847/1538-4357/aabae6},
archivePrefix = {arXiv},
       eprint = {1712.01527},
 primaryClass = {astro-ph.SR},
       adsurl = {https://ui.adsabs.harvard.edu/abs/2018ApJ...859...18P}
}

@ARTICLE{2018AJ....156...58B,
       author = {{Bailer-Jones}, C.~A.~L. and {Rybizki}, J. and {Fouesneau}, M. and {Mantelet}, G. and {Andrae}, R.},
        title = "{Estimating Distance from Parallaxes. IV. Distances to 1.33 Billion Stars in Gaia Data Release 2}",
      journal = {\aj},
         year = 2018,
        month = aug,
       volume = {156},
       number = {2},
          eid = {58},
        pages = {58},
          doi = {10.3847/1538-3881/aacb21},
archivePrefix = {arXiv},
       eprint = {1804.10121},
 primaryClass = {astro-ph.SR},
       adsurl = {https://ui.adsabs.harvard.edu/abs/2018AJ....156...58B}
}

\end{document}